\documentclass[twocolumn,aps,showpacs,prl]{revtex4}
\usepackage{graphicx}
\usepackage{epstopdf}
\usepackage{amsmath}
\usepackage{multirow}

\begin{document}

\title{First-principles study of pressure-induced magnetic phase transitions
in ternary iron selenide K$_{0.8}$Fe$_{1.6}$Se$_2$}

\author{Lei Chen$^{1,2}$}
\author{Xun-Wang Yan$^{1,2}$}
\author{Zhong-Yi Lu$^{1}$}\email{zlu@ruc.edu.cn}
\author{Tao Xiang$^{2,3}$}\email{txiang@iphy.ac.cn}

\date{\today}

\affiliation{$^{1}$Department of Physics, Renmin University of
China, Beijing 100872, China}

\affiliation{$^{2}$Institute of Theoretical Physics, Chinese Academy
of Sciences, Beijing 100190, China }

\affiliation{$^{3}$Institute of Physics, Chinese Academy of
Sciences, Beijing 100190, China }

\begin{abstract}

We have studied the pressure effect on electronic structures and magnetic orders of
ternary iron selenide K$_{0.8}$Fe$_{1.6}$Se$_2$ by the first-principles electronic
structure calculations. At low pressure, the compound is in the blocked checkerboard
antiferromagnetic (AFM) semiconducting phase, as observed by the neutron scatting
measurements. Applying pressure induces two phase transitions, first from the blocked
checkerboard AFM semiconducting phase to a collinear AFM metallic phase around 12 GPa,
and then to a non-magnetic metallic phase around 25 GPa, respectively. Our results help
to clarify the recent experimental measurements under pressure.

\end{abstract}


\maketitle

The discovery of iron-based superconductors\cite{kamihara} has stimulated great interest
on the investigation of unconventional superconducting mechanism. A system of particular
interest is the recently discovered potassium or other mono-valent element intercalated
ternary iron-selenide superconductors A$_y$Fe$_x$Se$_2$ (A=K, Rb, Cs, and/or
Tl)\cite{chen,Cs,fang}. These materials show rich magnetic phase diagrams and many
unusual physical properties that have not been found in other iron-based superconductors.
A thorough investigation may deepen our understanding on the  interplay between
superconducting and antiferromagnetic fluctuations.

AFe$_2$Se$_2$ has the ThCr$_{2}$Si$_{2}$ type crystal structure (Fig. 1a), isotructural
to the 122-type iron pnictides \cite{rotter}. However, from both experimental and
theoretical studies\cite{basca,bao1,yan11}, it was found that the chemically stable phase
of A$_y$Fe$_x$Se$_2$ is mainly contributed by A$_{0.8}$Fe$_{1.6}$Se$_2$ with a fivefold
expansion of the parent ThCr$_2$Si$_2$ unit cell in the $ab$ plane and a $\sqrt{5}\times
\sqrt{5}$ Fe vacancy order (Fig. 1b and 1c). Unlike the collinear \cite{cruz,ma1,yan} or
bi-collinear\cite{ma,bao,shi} antiferromagnetic (AFM) order found in the parent compounds
of other iron-based superconductors, the ground state of A$_{0.8}$Fe$_{1.6}$Se$_2$ was
found to have a novel blocked checkerboard AFM order with a band gap of 400-600 meV
\cite{yan11,cao} and a giant magnetic moment of 3.31 $\mu_B$/Fe formed below a Neel
temperature of 559$K$\cite{muSR,bao1}. In the blocked checkerboard AFM state, as shown in
Fig. 2(a), all four Fe moments within a tetramer are ordered in parallel, while the Fe
moments between two neighboring tetramers are ordered in anti-parallel.

The Fe vacancy order and the related distortion in A$_y$Fe$_x$Se$_2$ can be altered by
chemical doping or by applying pressure. This would in turn affect the stability of
different AFM orders in these materials. Unlike chemical doping, an advantage of applying
pressure is that it can tune or modify electronic structures and interactions of
materials without changing the chemical structures. This provides a unique opportunity to
clearly probe the microscopic origin of antiferromagnetic and superconducting pairing
interactions, as demonstrated in the study of other iron-based
superconductors\cite{pressure}. On the other hand, it is very difficult to experimentally
characterize the electronic structures and magnetic properties in high pressures by using
spectrum probing. This makes theoretical calculations substantially important to study
high pressure effect. The recent transport measurements already showed that increasing
pressure can suppress the superconducting transition temperature and lead to a
semiconducting-to-metallic phase transition at about 9 GPa for A$_y$Fe$_x$Se$_2$
\cite{zhao}. But no characterization information on the electronic structure and magnetic
order is available meanwhile.

\begin{figure}[t]
\includegraphics[width=8.0cm]{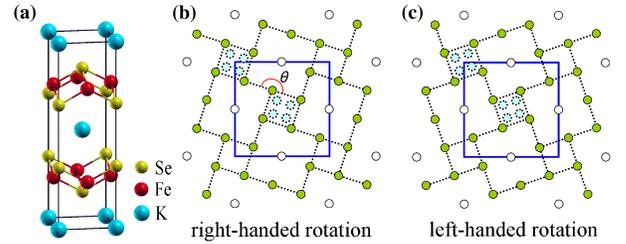}
\caption{(Color online) K$_y$Fe$_x$Se$_2$ with the ThCr$_{2}$Si$_{2}$-type structure: (a)
a tetragonal unit cell containing two formula units without any vacancy; (b) and (c) are
schematic top view of an Fe-Fe square layer with one-fifth Fe-vacancies ($x=1.6$) ordered
in $\sqrt{5}\times \sqrt{5}$ superstructure with the right- and left-chirality,
respectively. The squares enclosed by the solid lines denote the unit cells. The filled
circles denote the Fe atoms while the empty circles denote the Fe-vacancies. After the
tetramer lattice distortion, the four closest Fe atoms around a Se atom form a block
namely tretramer, represented by dashed (blue) circles in (b) and (c). } \label{figa}
\end{figure}

To investigate the pressure effect, we have performed first-principles electronic
structure calculations on K$_{0.8}$Fe$_{1.6}$Se$_2$ from 0 to 30 GPa. We find that
K$_{0.8}$Fe$_{1.6}$Se$_2$ undergoes two phase transitions at about 12 and 25 GPa,
respectively. The first transition is from the blocked checkerboard AFM semiconducting
state into the collinear AFM metallic state, and the second is from the collinear AFM
metallic state into a nonmagnetic metallic state. To our knowledge, such a
pressure-induced phase transition between two different magnetic phases has not been
found in other Fe-based superconductor.

In our calculations the plane wave basis method was used. We adopted the generalized
gradient approximation (GGA) with Perdew-Burke-Ernzerhof (PBE) formula \cite{pbe} for the
exchange-correlation potentials. The ultrasoft pseudopotentials \cite{vanderbilt} were
used to model the electron-ion interactions. The kinetic energy cut-off and the charge
density cut-off of the plane wave basis were chosen to be 800 eV and 6400 eV,
respectively. A mesh of $18\times 18\times 8$ k-points was sampled for the Brillouin-zone
integration while the Gaussian broadening technique was used in the cases of metallic
states. {\it The lattice parameters and the internal coordinates of all ions were fully
optimized by minimizing the enthalpy at a given pressure}.

In the calculations, we adopted a $\sqrt{5}\times\sqrt{5}\times 1$ tetragonal supercell,
which contains two FeSe layers with total 16 Fe atoms and 4 Fe-vacancies, 20 Se atoms,
and 8 K atoms and 2 K-vacancies, to represent K$_{0.8}$Fe$_{1.6}$Se$_2$. K atoms can have
two inequivalent positions in each supercell. However, from both theoretical
calculations\cite{yan11} and experimental measurements\cite{basca}, we know that the
electronic and magnetic structures of these materials are hardly affected by the
different configurations of K atoms.

From the previous calculations \cite{yan11}, we find that the tetramer lattice
distortion, of which the four Fe atoms within a tetramer shrink to form a compact square
driven by the chemical-bonding between Fe and Se, plays an essential role in stabilizing
the blocked checkerboard AFM state. Without such tetramer distortion, the energy of the
collinear AFM state (Fig. 2(b)) is in fact about 43 meV/Fe lower than that of the blocked
checkerboard AFM state. This collinear AFM state has also a semiconducting gap. But its
gap is only of dozen of meV, much smaller than in the blocked checkerboard AFM state. In
the collinear AFM state, each pair of next nearest Fe moments are AFM ordered. It
suggests that the Se-bridged superexchange AFM interaction dominates over other exchange
interactions, similar as in iron-pnictides.\cite{ma1}

\begin{figure}[tb]
\includegraphics[width=7.0cm]{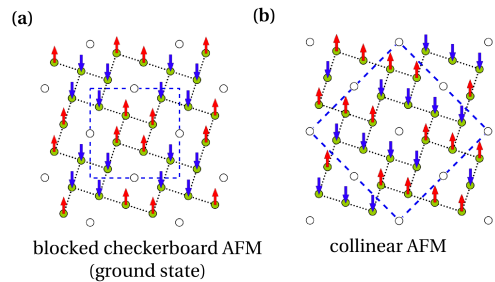}
\caption{(Color online) Schematic representation of the blocked checkerboard AFM state
(a) and the collinear AFM state (b) in each Fe layer. Fe vacancies are represented by
empty circles and Fe spins are shown by arrows. The dashed squares are the magnetic unit
cells. }
\end{figure}

To clarify the phase diagram, we have calculated the enthalpy $H$ under pressure. It
should be emphasized that it is the minimum of the enthalpy, rather than the minimum of
the total energy, that determines the stability of a phase under a pressure at zero
temperature. The enthalpy $H$ is defined by $H=E+pV$, where $E$, $p$, and $V$ stand for
the total energy (internal energy), pressure, and volume, respectively. In the
equilibrium, the pressure is thus equal to the minus derivative of the internal energy
with respect to the volume, $p=-dE/dV$. At the transition point between two phases, the
derivative of the total energy $E$ with respect to the volume $V$ should be equal for
these two phases. Thus the transition pressure can be also determined from the slope of
the common tangent lines on the $E$ versus $V$ curves of the two phases.

As there are Fe vacancies in A$_{0.8}$Fe$_{1.6}$Se$_2$, K atoms may enter these vacancy
positions under high pressures. We have carried out extensive calculations to examine
this possibility. However, we find that the enthalpy of the system with Fe vacancies
being partially or fully occupied by K atoms is always much higher than that of the
system without K occupation on Fe vacancies. Thus we need only to consider the system
with the $\sqrt{5} \times \sqrt{5}$-ordered Fe vacancy superstructure.

\begin{figure}[tb]
\includegraphics[width=7.5cm]{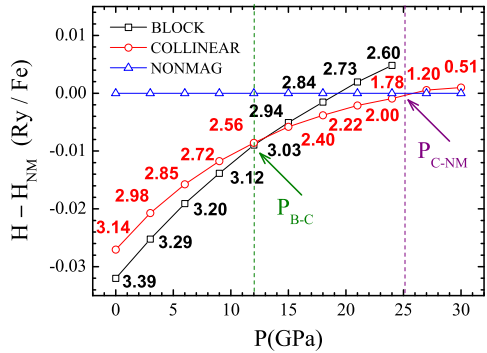}
\caption{(Color online) The enthalpy as a function of pressure for the blocked
checkerboard AFM, collinear AFM states, and non-magnetic states, respectively. The
enthalpy of the nonmagnetic state is taken as a reference. The number besides each point
is the magnetic moment for the corresponding magnetic state.} \label{fig3}
\end{figure}

Figure \ref{fig3} shows the pressure dependence of the enthalpy in the blocked
checkerboard AFM, collinear AFM, and non-magnetic states, respectively. Two phase
transitions are observed. In the low pressure regime, the system is in the semiconducting
blocked checkerboard AFM phase. At about 12 GPa, a transition from the blocked
checkerboard AFM to a metallic collinear AFM phase occurs. Between 12 and 25 GPa, the
collinear AFM phase has the smallest enthalpy. Above 25 GPa, the second transition occurs
and the system becomes non-magnetic. We have also calculated the enthalpies for the other
magnetic phases shown in Fig. 3 of Ref. [\onlinecite{yan11}] respectively, and found that
they are all substantially larger than the corresponding enthalpy of the collinear AFM
phase.

\begin{figure}
\includegraphics[width=7.5cm]{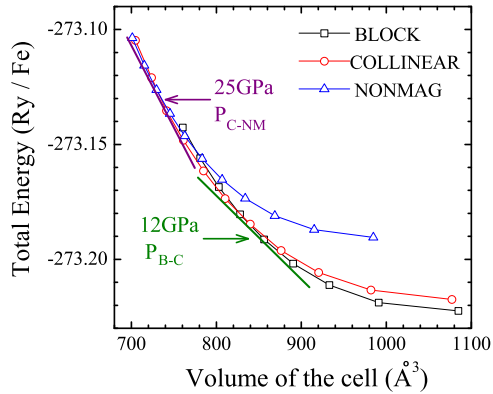}
\caption{(Color online) The total energy $E$ versus the volume $V$ for the blocked
checkerboard AFM, the collinear AFM, and the non-magnetic states, respectively. From the
slopes of the two common tangent lines, we find that the critical pressures are about 12
GPa and 25 GPa, respectively. They are consistent with the transition pressures
determined from the enthalpy, \textbf{P}$_{B-C}$ and \textbf{P}$_{C-NM}$, as shown in
Fig. \ref{fig3}} \label{fig4}
\end{figure}

As mentioned above, the phase transition pressures can be also determined from the slopes
of the common tangent lines in the total energy versus volume curves. Fig. \ref{fig4}
shows the volume dependence of the total energy. Here the total energy is calculated with
optimization at a fixed volume and the energy versus volume curves are given by Murnaghan
equations of state fitted to the calculated points. From the slopes of the common tangent
lines, we find that the two critical pressures are about 12 GPa and 25 GPa, respectively.
They agree with the values of critical pressures determined from the enthalpy.

\begin{figure}
\includegraphics[width=7.5cm]{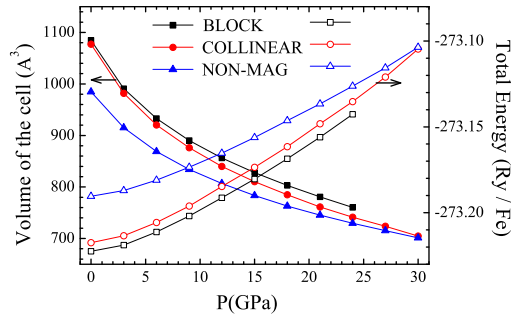}
\caption{(Color online) Pressure dependence of the volume (left) and the total energy
(right) for the blocked checkerboard AFM, the collinear AFM, and the non-magnetic states,
respectively.} \label{fig5}
\end{figure}

By further inspecting the pressure dependence of the volume and the total energy (Fig.
\ref{fig5}), we find that the cell volume in the collinear AFM phase drops faster with
pressure than in the blocked checkerboard AFM phase. However, the difference in the total
energy between these two phases remains almost unchanged under pressure. This is due to
the fact that with increasing pressure the collinear AFM state will become more and more
metallic, whereas the blocked checkerboard AFM state remains in an insulating phase.
Consequently, the enthalpy of the collinear AFM phase will become lower than that of the
blocked checkerboard AFM phase in high pressure.

By applying a pressure, the unit cell of K$_{0.8}$Fe$_{1.6}$Se$_2$ in the $ab$-plane is
shrunk in a similar way for both the collinear AFM and blocked checkerboard AFM phases.
The main difference of the pressure effect upon these two phases lies in the contraction
of the distance between Fe and Se atoms along the $c$-axis. The Fe-Se separation along
the $c$-axis in the collinear AFM phase is averagely about 2.40 \AA, 2.30 \AA, 2.23 \AA,
and 2.18 \AA~ at 0 GPa, 12 GPa, 24 GPa, and 27 GPa, respectively. The corresponding
separation in the blocked checkerboard AFM phase is 2.44 \AA,  2.36 \AA, and 2.30 \AA~ at
0 GPa, 12 GPa, and 24 GPa, respectively. Thus the contraction along the $c$-axis with the
pressure is more pronounced in the collinear AFM phase than in the blocked checkerboard
AFM phase. It is known that the magnetic moment of Fe is sensitive to the distance
between Fe and As or Se\cite{mazin}, thus the magnetic moment should become smaller with
increasing pressure. This is consistent with the result shown in Fig. \ref{fig3}. At zero
pressure, both the collinear AFM and blocked checkerboard AFM states are semiconducting,
the charge fluctuation is small and each Fe acquires a giant magnetic moment. In high
pressures, the collinear AFM state becomes metallic and the charge fluctuation between
the conduction and valence bands reduces significantly the effective moment.

In the collinear AFM state at 12 GPa, one can flip some of the Fe spins $\vec{S}$ to form
different magnetic states. From the energy differences among these states, we have
estimated the effective spin exchange interactions and found the nearest neighbor spin
exchange constant $J_1=-9~meV/S^2$ and the next-nearest neighbor exchange constant
$J_2=22~meV/S^2$, similar as in iron-pnictides\cite{ma1}.

\begin{figure}
\includegraphics[width=8.0cm]{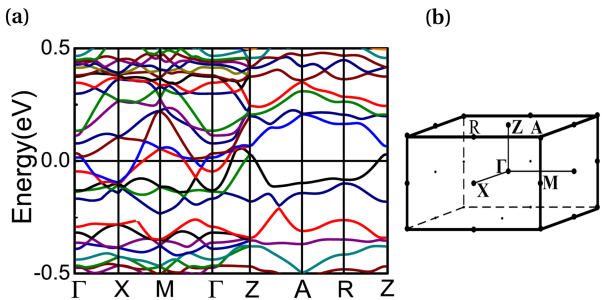}
\caption{(Color online) (a) Electronic band structure of K$_{0.8}$Fe$_{1.6}$Se$_2$ in the
collinear antiferromagnetic order (Fig. 2(b)) at a pressure of 12 GPa; (b) Brillouin
zone. Here the Fermi energy is set to zero. }
\label{fig6}
\end{figure}

Figure \ref{fig6} shows the electronic band structure of K$_{0.8}$Fe$_{1.6}$Se$_2$ in the
collinear AFM phase at 12 GPa. At this pressure, K$_{0.8}$Fe$_{1.6}$Se$_2$ is
semi-metallic. It contains both electron-type and hole-type charge carries. From our
previous study, we know that at the ambient pressure the collinear AFM state is a
meta-stable semiconductor with a band gap of 73 meV \cite{yan11}. High pressure turns it
into a metal. In contrast, K$_{0.8}$Fe$_{1.6}$Se$_2$ in the blocked checkerboard AFM
phase is always semiconducting with a band gap varying from 600 meV to 300 meV with the
increase of pressure up to 24 GPa.

The semiconductor-to-metal transition we find at about 12 GPa between the blocked
checkerboard AFM and the collinear AFM phases agrees qualitatively with the experimental
observation\cite{zhao}. But the transition pressure we obtained is slightly higher than
the experimental value (about 9 GPa) for K$_y$Fe$_x$Se$_2$. The difference may result
from two reasons. First, the samples used in experimental measurements are not in the
perfect K$_{0.8}$Fe$_{1.6}$Se$_2$ composition. They always contain extra Fe and K ions.
These extra ions will introduce itinerant charge carriers to the system and reduce the
value of the transition pressure. Second, as in all the density functional calculation
based on the PBE approximation for iron-pnicnides and iron-chalcogenides, the magnetic
moment of Fe is over-estimated in the collinear AFM metallic states\cite{ma1}. This means
that the charge fluctuation in the collinear AFM state is underestimated, which in turn
may result an overestimation for the critical pressure.

In conclusion, we have studied the pressure effect on K$_{0.8}$Fe$_{1.6}$Se$_2$ by
performing the first principles electronic structure calculations. Without external
pressure, the ground state is in the blocked checkerboard AFM semiconducting phase. By
applying a pressure, it undergoes two successive phase transitions, first to the
collinear AFM metallic phase at about 12 GPa, and then to the non-magnetic metallic phase
at about 25 GPa. Our result agrees qualitatively with the experimental measurement. It
suggests that the  semiconductor-to-metal phase transition observed by experiments is a
transition from the blocked checkerboard AFM phase to the collinear AFM metallic phase.
The metallic collinear AFM phase in high pressures shows both electron- and hole-type
Fermi surfaces, similar as in iron-pnictides or in FeSe. In analogy to BaFe$_2$As$_2$ or
other iron-based superconductors, we believe that the collinear AFM metallic phase of
K$_{0.8}$Fe$_{1.6}$Se$_2$ may become superconducting upon pressure and/or electron or
hole doping.

We would like to thank Professor Liling Sun for helpful discussions. This work is
partially supported by National Natural Science Foundation of China and by National
Program for Basic Research of MOST, China.

{\it Note added.} While in the preparation of this manuscript, we learnt of a paper by C.
Cao, M. Fang, and J. Dai on pressure effect on K$_y$Fe$_x$Se$_2$\cite{cao11}. Unlike what
we have done here, they used the total energy rather than the enthalpy to judge the
stability of a phase under pressure. Based on such a criterion, they claimed that a
so-called Neel-FM state is the most stable in the regime corresponding to the collinear
AFM phase in Fig. \ref{fig3}. We have subsequently calculated the enthalpy for this
Neel-FM state. But we find that its enthalpy is substantially higher than that of the
collinear AFM state up to 25 GPa.

\end{document}